\documentclass{article}

\usepackage[preprint]{neurips_2025}

\usepackage[utf8]{inputenc} % allow utf-8 input
\usepackage[T1]{fontenc}    % use 8-bit T1 fonts
\usepackage{hyperref}       % hyperlinks
\usepackage{url}            % simple URL typesetting
\usepackage{booktabs}       % professional-quality tables
\usepackage{amsfonts}       % blackboard math symbols
\usepackage{nicefrac}       % compact symbols for 1/2, etc.
\usepackage{microtype}      % microtypography
\usepackage{xcolor}         % colors

\usepackage{amsmath}
\usepackage{amssymb}
\usepackage{mathtools}
\usepackage{amsthm}
\usepackage{multirow}
\usepackage{wrapfig}
\usepackage{diagbox}

\RequirePackage{times}    % Integrate Times for here
\RequirePackage{xspace}
\RequirePackage[dvipsnames]{xcolor}
\RequirePackage{graphicx}
\RequirePackage{amsmath}
\RequirePackage{amssymb}
\RequirePackage{booktabs}
\usepackage{bm}
\definecolor{bronze}{rgb}{1,1,0.6}
\definecolor{silve}{rgb}{0.969,0.796,0.600}
\definecolor{gold}{rgb}{0.941,0.592,0.600}

\usepackage{amsthm}

\usepackage{algorithm}
\usepackage{algpseudocode}
\usepackage{cuted}
\usepackage{float}
\usepackage{caption}
\usepackage{comment}
\title{Deep Compositional Phase Diffusion for Long Motion Sequence Generation}  

\author{%
  Ho Yin Au\\
  Hong Kong Baptist University\\
  \texttt{cshyau@comp.hkbu.edu.hk} \\
  \And
  Jie Chen\thanks{Corresponding Author}\\
  Hong Kong Baptist University\\
  \texttt{chenjie@comp.hkbu.edu.hk} \\
  \And
  Junkun Jiang\\
  Hong Kong Baptist University\\
  \texttt{csjkjiang@comp.hkbu.edu.hk} \\
  \And
  Jingyu Xiang\\
  Hong Kong Baptist University\\
  \texttt{csjyxiang@comp.hkbu.edu.hk} \\
}

\begin{document}

\maketitle
\begin{figure}[ht]
\vspace{3mm}
\includegraphics[width=\textwidth]{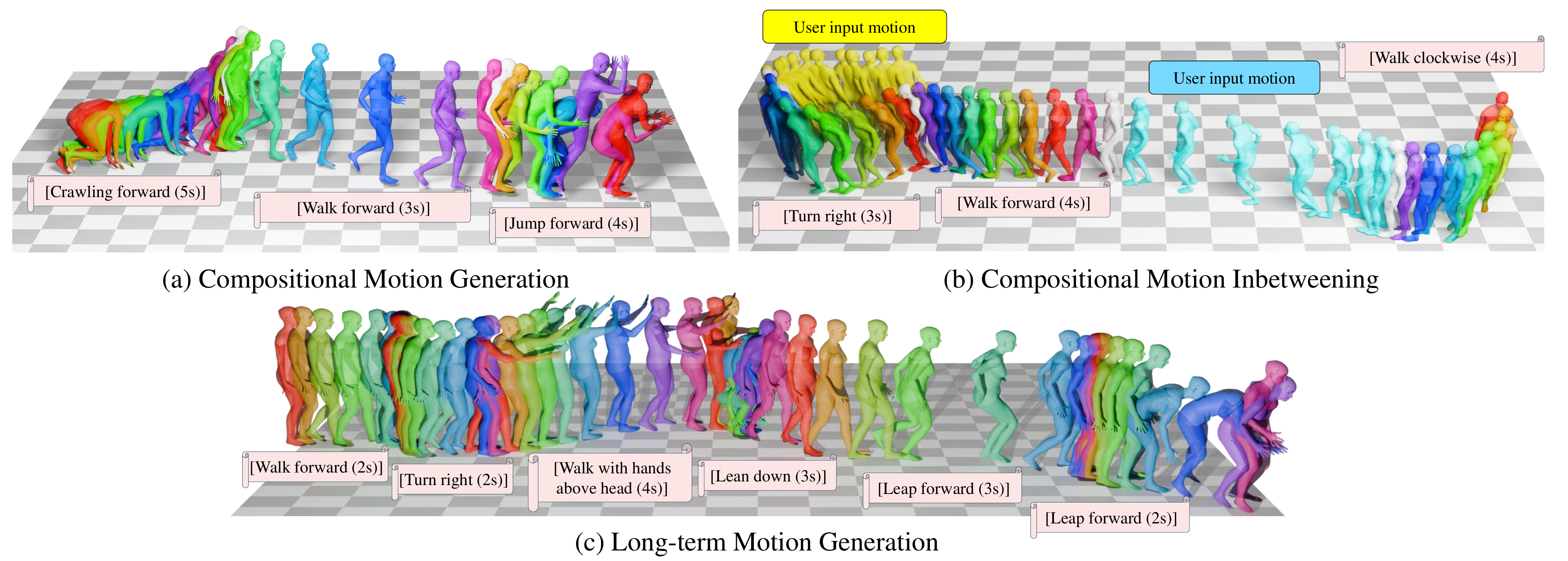}
\caption{Our Compositional Phase Diffusion framework produces high-quality composite motion sequences with smooth transitions and semantic alignment. (a) Compositional generation involves synthesizing multiple motion segments of varying lengths simultaneously, ensuring smooth transitions between segments. (b) Motion inbetweening allows users to select segments (in blue and yellow) and create conditional or unconditional bridging motions. (c) Long-term motion generation is achieved by scaling the framework with additional modules, enabling the parallel denoising of a larger number of motion segments. The rainbow color indicates time progression.}
\label{fig:teaser}
\vspace{3mm}
\end{figure}

\begin{abstract}
Recent research on motion generation has shown significant progress in generating semantically aligned motion with singular semantics. However, when employing these models to create composite sequences containing multiple semantically generated motion clips, they often struggle to preserve the continuity of motion dynamics at the transition boundaries between clips, resulting in awkward transitions and abrupt artifacts. To address these challenges, we present Compositional Phase Diffusion, which leverages the Semantic Phase Diffusion Module (SPDM) and Transitional Phase Diffusion Module (TPDM) to progressively incorporate semantic guidance and phase details from adjacent motion clips into the diffusion process. Specifically, SPDM and TPDM operate within the latent motion frequency domain established by the pre-trained Action-Centric Motion Phase Autoencoder (ACT-PAE). This allows them to learn semantically important and transition-aware phase information from variable-length motion clips during training. Experimental results demonstrate the competitive performance of our proposed framework in generating compositional motion sequences that align semantically with the input conditions, while preserving phase transitional continuity between preceding and succeeding motion clips. Additionally, motion inbetweening task is made possible by keeping the phase parameter of the input motion sequences fixed throughout the diffusion process, showcasing the potential for extending the proposed framework to accommodate various application scenarios. Codes are available at \textcolor{magenta}{https://github.com/asdryau/TransPhase}.

\end{abstract}
%discretization-invariant 

\section{Introduction}\label{sec:introduction}
% The technical challenge of compositional motion generation involves producing smooth transitions to connect motion pairs generated from text or user input.

Deep learning-based human motion generation holds significant potential for creating virtual humanoid animations and enhancing robotics applications. 
With more advanced modeling techniques~\cite{vaswani2017transformer, ho2020ddpm,song2020ddim} and more motion data being captured~\cite{punnakkal2021babel,jiang2022dual, guo2022t2m,jiang2025every}, motion generation models are evolving rapidly and can be adapted to a variety of multimodal generation tasks.
For example, text-to-motion generation allows animators to produce character animations by specifying semantic contexts using text prompts.
Current motion generation models~\cite{tevet2023mdm,guo2022t2m,guo2022tm2t,kim2023flame, zhang2024motiondiffuse} handle variable-length motion segments with singular semantics. However, when these models are applied to long-term compositional motion generation tasks, where they must generate $K$ motion segments sequentially for $K$ instructions, they often struggle with smooth transitions between segments.
Recently, there has been a growing focus on long-term compositional generation tasks~\cite{athanasiou2022teach,shafir2024priorMDM,yang2023pcmdm}, driven by the availability of BABEL-TEACH~\cite{punnakkal2021babel, athanasiou2022teach}, a dataset with text annotations for pairs of temporally connected motion segments, which aids motion models in learning transitions. Methods such as priorMDM~\cite{shafir2024priorMDM} use the learned transition knowledge to create transitional segments that smooth out pose differences between those generated by MDM~\cite{tevet2023mdm}.
However, these approaches often overlook the intrinsic kinematics of each segment, resulting in artifacts like over-smoothing or abrupt stops in transitions.

To generate motion clips aligned with specific semantic contexts and ensure smooth transitions, we introduce the Compositional Phase Diffusion framework. This framework simultaneously creates multiple motion clips from sequential semantic instructions, using denoised information from adjacent clips to enhance transition compatibility.
The Compositional Phase Diffusion framework consists of three main components: Action-Centric Periodic Autoencoder (ACT-PAE), Semantic Phase Diffusion Module (SPDM), and Transitional Phase Diffusion Module (TPDM). ACT-PAE, which builds upon DeepPhase~\cite{starke2022deepphase}, encodes each variable-length motion segment into a unified phase manifold. SPDM and TPDM then iteratively denoise phase parameters by incorporating semantic instructions and neighbouring phase signals. This approach effectively models the intrinsic dynamics of each motion segment within the latent motion frequency domain, ensuring both semantic alignment and smooth transitions. Additionally, the framework is scalable, allowing for an arbitrary number of modules to denoise multiple motion segments in parallel. This capability highlights its flexibility and scalability in generating motion sequences of varying lengths and facilitating motion inbetweening tasks.

Extensive experimental results demonstrate that the Compositional Phase Diffusion framework excels in both long-term compositional motion generation and motion inbetweening tasks, attributed to the semantic and transition-aware diffusion process. The key contributions are as follows:
\begin{itemize}
\item We introduce the Compositional Phase Diffusion framework, a scalable and efficient solution for various motion generation tasks. This framework can process an arbitrary number of motion segments of varying lengths simultaneously by leveraging parallel module execution, ensuring smooth and coherent transitions between clips.
\item Our framework incorporates three key components: the Action-Centric Periodic Autoencoder (ACT-PAE), the Semantic Phase Diffusion Module (SPDM), and the Transitional Phase Diffusion Module (TPDM). By operating within a unified phase latent space established by ACT-PAE, SPDM and TPDM collaboratively denoise motion representations while preserving semantic phase information and aligning transitional dynamics.
\item Extensive experiments validate the effectiveness of our framework, demonstrating significant improvements in long-term compositional motion generation and motion inbetweening tasks, showcasing its ability to produce high-quality, contextually relevant animations.
\end{itemize}

% \begin{itemize}
% \item We propose the Action-Centric Periodic Autoencoder (ACT-PAE) which advances the autoencoder from DeepPhase~\cite{starke2022deepphase} from fixed sliding window design to action-centric design with variable window length. ACT-PAE can capture action-specific dynamics in motion clips more efficiently, enhancing performance in subsequent text-to-motion tasks. 
% \item Building upon ACT-PAE, we introduce the Semantic Phase Diffusion Module (SPDM) and Transitional Phase Diffusion Module (TPDM). These modules are designed to inject semantic guidance and neighboring phase information into the motion diffusion process.
% % Specifically, TPDM leverage the local motion phase information of the preceding and succeeding motions in the diffusion process to facilitate phase consistency during transitions. 
% \item We investigated a highly efficient diffusion pipeline applicable for both long-term text-to-motion generation and motion inbetweening tasks. Specifically, multiple motion clips will be simultaneously denoised to ensure both the semantic alignment and phase consistency of the overall compositional motion sequence.
% \end{itemize}

\section{Related work}
\vspace{-2mm}

\noindent\textbf{Motion Phase Modeling.}
% \subsection{Motion Phase Modeling}
Pioneering approaches~\cite{holden2017PFNN, starke2020localphase, mason2022stylelocalphase} incorporate explicit phase inputs, such as foot contact during walking, to achieve smooth motion extrapolation and transition.
DeepPhase~\cite{starke2022deepphase} further extends this concept by developing a Periodic Autoencoder (PAE) that encodes motion segments into phase latent parameters, i.e., frequency ($\mathbf{F}$), amplitude ($\mathbf{A}$), offset ($\mathbf{B}$), and phase shift ($\mathbf{S}$). These parameters help generate periodic motion patterns and smooth transitions, minimizing artifacts like over‐smoothing and sudden stops.
Building upon PAE, PhaseBetweener~\cite{starke2023MIBPhase} and RSMT~\cite{tang2023RSMT} tackle motion inbetweening tasks by autoregressively generating motion frames and phase parameters. 
Meanwhile, DiffusionPhase~\cite{wan2023diffusionphase} adopts the MLD~\cite{chen2023mld} framework to denoise the periodic latents based on input text and conditioned pose. However, the fixed-length convolution scheme of PAE leads to instability in training objectives, as variable-length motions are encoded into a varying number of phase latent codes.

\noindent\textbf{Text-to-Motion Generation.}
% \subsection{Text-to-Motion Generation}
Several methods have been utilized diffusion models~\cite{ho2020ddpm,song2020ddim} with a single text prompt, including MDM~\cite{tevet2023mdm}, MLD~\cite{chen2023mld}, MotionDiffuse~\cite{zhang2024motiondiffuse}, and DiffusionPhase~\cite{wan2023diffusionphase}. Notably, MDM~\cite{tevet2023mdm} applies the Diffusion Model to raw pose sequences conditioned on text encoded by CLIP~\cite{radford2021clip}. 
Building upon MDM, PhysDiff~\cite{yuan2023physdiff} and GMD~\cite{karunratanakul2023gmd} have been developed to enhance physical plausibility and trajectory control in the generated motion. However, due to the limited segment lengths of datasets like HumanML3D~\cite{guo2022t2m} and BABEL~\cite{punnakkal2021babel}, with maximum frames of 196 and 250, respectively, these models struggle to generate longer motion sequences.

\noindent\textbf{Learning-based Motion Inbetweening}
% \subsection{Learning-based Motion Inbetweening}
is achieved through two main approaches. 1) Autoregressive frame generation: Motion frames are sequentially generated to connect segment boundaries~\cite{harvey2018RTN, harvey2020RMIB}. Methods like DiffusionPhase~\cite{starke2023MIBPhase} and RSMT~\cite{tang2023RSMT} further incorporate motion phase modeling for smoother, phase-aware transitions. 2) Segment interval infilling: Transitional segments of specified length are created to bridge segment boundaries~\cite{kaufmann2020CMIF, hernandez2019STinpaint}. Methods like CMB~\cite{kim2022CMIB} and MDM~\cite{tevet2023mdm} extend this by integrating semantic conditions into the inbetweening motion generation process.

\noindent\textbf{Long Motion Sequence Generation}
% \subsection{Long Motion Sequence Generation}
can be approached in two ways: sequential generation and parallel generation. Sequential generation methods such as TEACH~\cite{athanasiou2022teach}, PCMDM~\cite{yang2023pcmdm}, M2D2M~\cite{chi2024m2d2m}, and InfiniMotion~\cite{zhang2024infinimotion} generate motion segments one after another in an autoregressive manner. Analogous to traditional motion graph based approaches~\cite{kovar2008motiongraph, min2012motiongraphpp, au2022choreograph, au2024rechoreonet}, these methods require that the generated segments not only align with the current input semantics, but also transition smoothly from previously generated segments.
For parallel generation, priorMDM~\cite{shafir2024priorMDM} generates semantic motion segments independently and then synthesizes blending transitional segments using a diffusion model. Note that the frameworks above typically model transitions in the raw motion space, which may lead to slight discontinuities at the segment boundaries. To address this, motion inbetweening techniques are usually employed to smooth the transition boundaries. For example, TEACH~\cite{athanasiou2022teach} uses spherical linear interpolation (SLERP) to create motion frames connecting boundary poses.

\vspace{-1mm}
\section{Compositional Phase Diffusion}
\vspace{-2mm}
We propose three key components for the framework: the Action-Centric Periodic Autoencoder (ACT-PAE), the Transitional Phase Diffusion Module (TPDM), and the Semantic Phase Diffusion Module (SPDM). ACT‐PAE creates a motion latent manifold that captures important semantic and transition‐aware phase information for each motion segment $\mathbf{X} \in \mathbb{R}^{N \times E}$ and represent them as a set of latent variables $\mathbf{P} = [\mathbf{F},\mathbf{A},\mathbf{B},\mathbf{S}]$. Leveraging such ACT‐PAE latent space, TPDMs refine phase latents of the current segment using the \textbf{phase dynamics information from adjacent motions}, while SPDM incorporates \textbf{semantic information} into the diffusion process. Details of these components will be covered in Sec.~\ref{sec:components}. 

With these innovative elements, we adapt the Compositional Phase Diffusion framework to various motion generation tasks. For the compositional motion generation and motion inbetweening tasks, SPDMs and TPDMs gradually integrate semantic information and phase dynamics information from adjacent segments throughout the denoising process of sequentially connected segments.
For the long-term motion generation task, the phase dynamics of a motion segment will progressively propagate bidirectionally along the timeline during the denoising process. This promotes mutual phase dynamics adjustment between segments, increases their transition-awareness, and thereby enhances overall motion consistency.
By blending a series of transition-aware motion segments, we create a cohesive motion sequence composed of a series of semantically meaningful segments and seamless transitions in between. Further details are provided in Sec.~\ref{sec:framework_extensions}.

\begin{figure*}[t]
\centering
\includegraphics[width=0.98\linewidth]{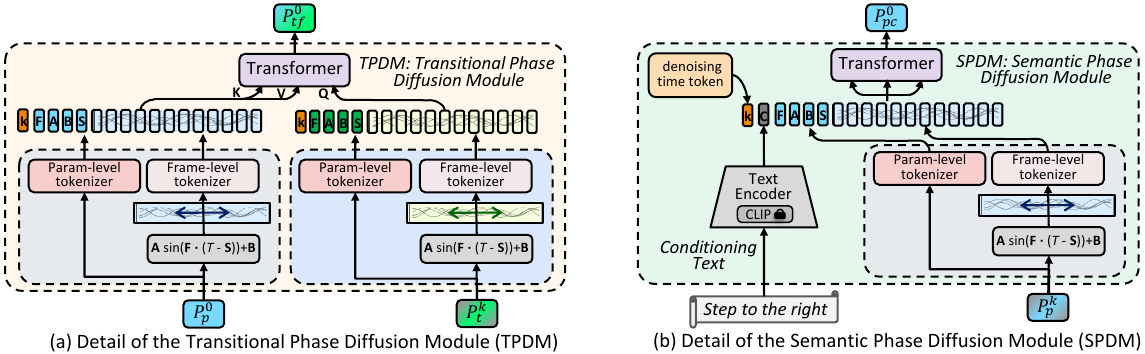}
\caption{Module detail of TPDM and SPDM. (a) The TPDM uses the clean phase latents from the adjacent segment $\mathbf{P}_\mathbf{p}^0$ to denoise the current motion phase $\mathbf{P}_\mathbf{t}^k$, making the current denoising motion align with the motion dynamics of the adjacent motion segment. (b) The SPDM utilizes the text embedding $C$ from CLIP to denoise the motion phase parameters of the current denoising motion.
}
\vspace{-0.2cm}
\label{fig:architecture}
\end{figure*}

\subsection{Key Components}\label{sec:components}

\subsubsection{ACT-PAE: Action-Centric Periodic Autoencoder} \label{sec:actpae}
Our ACT-PAE builds upon the transformer-based motion autoencoder architecture from ACTOR~\cite{petrovich2021actor}.
ACT-PAE encoder first processes input motion $\mathbf{X} \in \mathbb{R}^{N \times E}$ of $N$ frames into four phase parameters $\mathbf{F},\mathbf{A},\mathbf{B},\mathbf{S} \in \mathbb{R}^{Q}$. Unlike PAE~\cite{starke2022deepphase}, which processes motion using a convolution scheme and derives phase parameters from the FFT results, the ACT-PAE encoder directly processes variable-length motion using a transformer and predicts their phase parameters. To enforce latent space periodicity, these parameters are parameterized into a periodic signal $\mathbf{Q} \in \mathbb{R}^{N \times Q}$ using the following equation:
\begin{equation}\label{eq:phase_reparam}
\mathbf{Q} = \mathbf{A} \sin (\mathbf{F} \cdot (T-\mathbf{S}))+\mathbf{B}.
\end{equation}
Here $T$ represents the time difference of each frame in the motion \textit{relative to the center} of the motion segment. The $\sin$ function parameterizes $\mathbf{F} \cdot (T-\mathbf{S})$ into a periodic sine wave, transforming frequency and phase shift information into a sinusoidal basis. This representation allows ACT-PAE to capture the underlying phase dynamics of the motion effectively.
Finally, the ACT-PAE decoder takes $\mathbf{Q}$ to predict the motion $\mathbf{\hat{X}}$. The entire ACT‐PAE is trained with L2 loss.

% The key advantage of ACT-PAE lies in its ability to process semantically meaningful motion (e.g., $\mathbf{X_p}$ and $\mathbf{X_s}$) as a unified entity.
The main advantage of ACT-PAE lies in its ability to capture unified phase dynamics within semantically meaningful motion (e.g., $\mathbf{X_p}$ and $\mathbf{X_s}$). Its architecture handles variable-length motion input, which eliminates the need for fixed-window motion slicing and thus preserves complete semantic and transition-aware phase information in phase latents. By doing so, ACT‐PAE standardizes the training objective for the subsequent motion diffusion modules more effectively than the fixed-window process used in PAE, which results in an undetermined number of phase latents. Importantly, we have changed the sinusoidal positional embedding module $PE$ and the time window $T$ to accommodate variable‐length motion encoding. For example, $T$ can be parameterized for normalized action progression ($-1$ to $1$ across $N$ frames), or actual time duration ($-\frac{N}{2}$ to $\frac{N}{2}$ across $N$ frames). Details of $PE$ and $T$ adjustments will be provided in the Appendix.

\subsubsection{SPDM: Semantic Phase Diffusion Module} \label{sec:spdm}
SPDM is designed to denoise phase parameters so that the corresponding decoded motion segment is aligned to the semantic condition. In text-to-motion settings, SPDM employs the pre-trained \textit{CLIP-ViT-B/32}~\cite{radford2021clip} to encode the input text conditions into embedding vector $C_\mathbf{p}$, as shown in Fig.~\ref{fig:architecture}(b). This embedding guides the denoising process of the phase parameters $\mathbf{P}_\mathbf{p}^k$, which are encoded by ACT‐PAE, for the semantically conditioned motion $\mathbf{X_p}$ via $\mathbf{P}_{\mathbf{p}c}^0 = \mathcal{F}_{\text{S}}(k, C_\mathbf{p}, \mathbf{P}_\mathbf{p}^k)$. Here, $k$ indicates the denoising time step. Note that the input phase parameters $\mathbf{P}$ are parameterized as both \textit{param-level tokens} $[\mathbf{F},\mathbf{A},\mathbf{B},\mathbf{S}]$ and \textit{frame-level tokens} which constitute the periodic signal $\mathbf{Q}$ created using Equation~\ref{eq:phase_reparam}. These frame-level tokens explicitly outline the spatio-temporal motion context in the phase parameters, assisting SPDM in monitoring the current semantic context during the phase parameter denoising process. Finally, a self-attention transformer~\cite{vaswani2017transformer} is employed to derive semantically-denoised parameters $\mathbf{P}_\mathbf{pc}^0$. 

\begin{figure*}[t]
\centering
\includegraphics[width=0.98\linewidth]{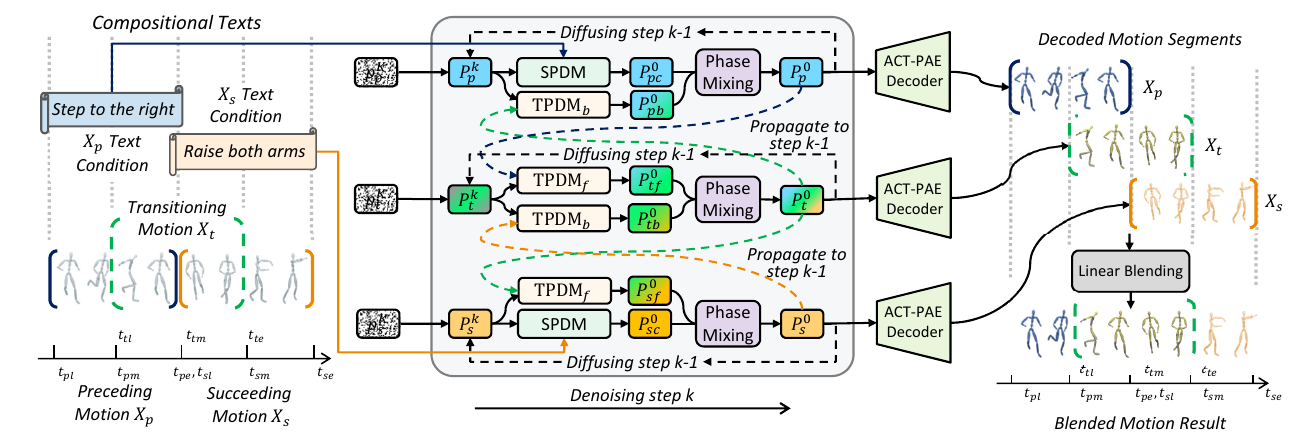}
\vspace{-0.3cm}
\caption{Illustration of the phase diffusion pipeline for the compositional motion generation task. SPDMs and TPDMs guide the denoising of motion segments through \textbf{semantic information} and the \textbf{phase dynamics information from adjacent motions}, respectively. The denoised results are combined via phase mixing and either diffused back to step $k-1$ or fed into the ACT-PAE decoder at the final step to produce motion segments, which are then linearly blended to create a long sequence.}
\vspace{-0.3cm}
\label{fig:framework}
\end{figure*}

\subsubsection{TPDM: Transitional Phase Diffusion Module} \label{sec:tpdm}
TDPM is designed to denoise phase parameters such that the resulting decoded motions are transitionally aligned with adjacent motions. Depending on the specific application scenario, these adjacent motions may come from either the forward or backward direction, which will be explained in Sec.~\ref{sec:framework_extensions}.

Fig.~\ref{fig:architecture}(a) illustrates the TPDM architecture, which leverages the clean phase parameters of the forward, preceding motion $\mathbf{P}_\mathbf{p}^0$ to denoise the phase parameters of the transitioning motion $\mathbf{P}_\mathbf{t}^k$, $\mathbf{P}_{\mathbf{t}f}^0 = \mathcal{F}_{\text{T}_f}(k, \mathbf{P}_\mathbf{t}^k, \mathbf{P}_\mathbf{p}^0)$. $f$ indicates that the denoising process is conditioned by the phase dynamics of the forward, preceding motion.
Similar to the SPDM design, both \textit{param-level tokens} and \textit{frame-level tokens} are computed for $\mathbf{P}_\mathbf{p}^k$. The motion context provided by the frame-level tokens assists the TPDM in ensuring alignment of the motion dynamics during the denoising process. Finally, a cross-attention transformer~\cite{vaswani2017transformer} processes all the $\mathbf{P}_\mathbf{t}^k$ and $\mathbf{P}_\mathbf{p}^0$ tokens to predict  phase noise $\mathbf{P}_{\mathbf{t}f}^0$.

As can be seen in Fig.~\ref{fig:framework}, there are at least two TPDM modules involved in compositional motion generation: $\text{TPDM}_f$ utilizes preceding motion phase $\mathbf{P}_\mathbf{p}^0$ to denoise $\mathbf{P}_\mathbf{t}^k$, $\mathbf{P}_{\mathbf{t}f}^0 = \mathcal{F}_{\text{T}_f}(k, \mathbf{P}_\mathbf{t}^k, \mathbf{P}_\mathbf{p}^0)$ and $\text{TPDM}_b$ utilizes succeeding motion phase $\mathbf{P}_\mathbf{s}^0$ instead: $\mathbf{P}_{\mathbf{t}b}^0 = \mathcal{F}_{\text{T}_b}(k, \mathbf{P}_\mathbf{t}^k, \mathbf{P}_\mathbf{s}^0)$. 
These two modules work together to ensure that dynamic coherence is maintained in both \textit{forward} and \textit{backward} directions throughout the composited long sequence.

SPDM and TPDMs are implemented as $\epsilon$-models, and their training procedures follow those of traditional diffusion frameworks~\cite{karunratanakul2023gmd, song2020ddim}. Details for SPDM and TPDM are provided in the Appendix.

\subsection{Applications}\label{sec:framework_extensions}
\subsubsection{Compositional Motion Pair Generation}\label{sec:compositional}

The compositional motion pair generation task focuses on creating two sequentially connected motion segments, $\mathbf{X_p}$ and $\mathbf{X_s}$.
To ensure a smooth transition while maintaining semantic alignment, we develop a compositional motion diffusion pipeline that progressively incorporates the \textbf{semantic information} and the \textbf{phase dynamics information from adjacent segments} in the diffusion process. This phase dynamics information exchange enhances phase alignment between $\mathbf{X_p}$ and $\mathbf{X_s}$, and facilitates the creation of an intermediate transition segment $\mathbf{X_t}$\footnote{We define the concept of the transition motion $\mathbf{X_t}$ to be the segment covering exactly the second half of $\mathbf{X_p}$ and the first half of $\mathbf{X_s}$.}, which is linearly blended into the output to further smooth the segment boundary.

The pipeline detail is shown in Fig.~\ref{fig:framework} and described in Algorithm~\ref{alg:diffusion} in the Appendix. 
During the denoising step $k$, SPDM semantically denoises the phase latents $\mathbf{P}_\mathbf{p}^k$ and $\mathbf{P}_\mathbf{s}^k$ for $\mathbf{X_p}$ and $\mathbf{X_s}$ based on their respective semantic conditions. 
$\text{TPDM}_f$ and $\text{TPDM}_b$ then estimate $\mathbf{P}_{\mathbf{p}b}^0$, $\mathbf{P}_{\mathbf{t}f}^0$, $\mathbf{P}_{\mathbf{t}b}^0$ and $\mathbf{P}_{\mathbf{s}f}^0$ by combining and mixing information from temporally adjacent phase latents (i.e., $\mathbf{P}_\mathbf{p}^0$, $\mathbf{P}_\mathbf{t}^0$, and $\mathbf{P}_\mathbf{s}^0$) from the earlier denoising step $k+1$. 
For instance, $\mathbf{P}_\mathbf{t}^k$ is denoised with $\mathbf{P}_\mathbf{p}^0$ and $\mathbf{P}_\mathbf{s}^0$ from step $k+1$, while the resulting $\mathbf{P}_\mathbf{t}^0$ from step $k$ helps denoising $\mathbf{P}_\mathbf{p}^{k-1}$ and $\mathbf{P}_\mathbf{s}^{k-1}$ in step $k-1$, demonstrating the exchange of phase dynamics throughout diffusion process. 

\begin{figure*}[t]
\centering
\includegraphics[width=1\linewidth]{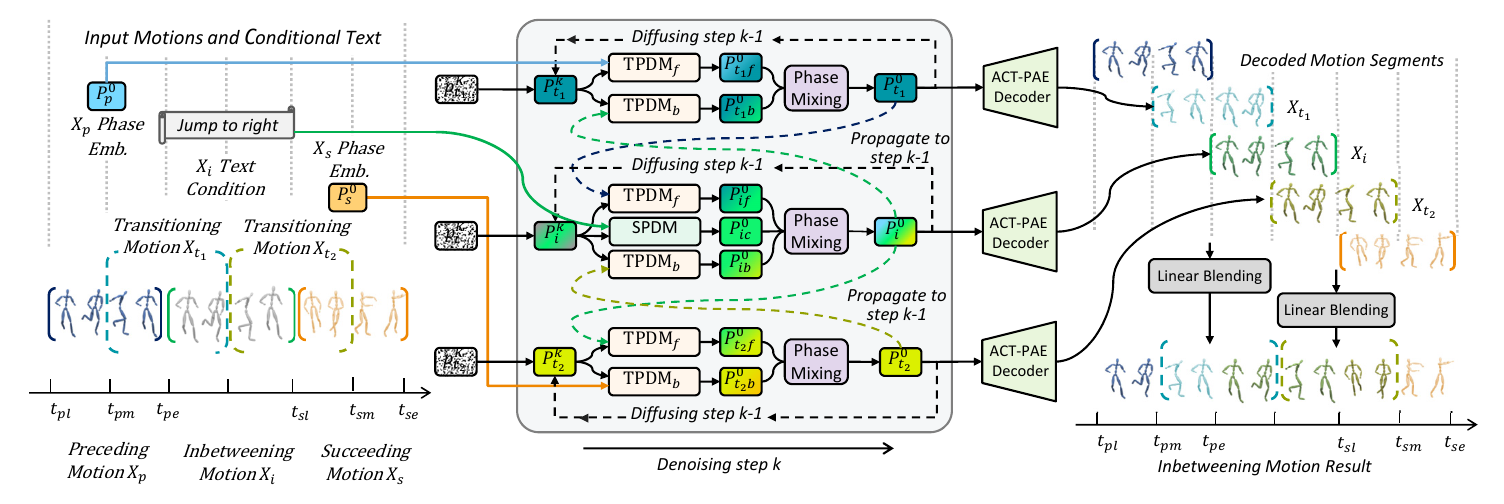}
\vspace{-0.7cm}
\caption{
The phase diffusion pipeline for the motion inbetweening tasks. The inbetweening motion $\mathbf{X_i}$ along with two transitional motions $\mathbf{X_{t_1}}$ and $\mathbf{X_{t_2}}$ are simultaneously generated. These motions are subsequently linear blended to form the final inbetweening output.
}
\vspace{-0.3cm}
\label{fig:inference_pipeline}
\end{figure*}

The predictions from both SPDM and TPDM are then combined using the Phase Mixing Equation:
\begin{equation}\label{eq:phase_mix}
\mathbf{P}_\cdot^0 = r \frac{\mathbf{P}_{\mathbf{\cdot}f}^0 + \mathbf{P}_{\mathbf{\cdot}b}^0}{2} + (1-r) \mathbf{P}_{\mathbf{\cdot}c}^0
\end{equation}
The variable $r$ is defined as $r=(\frac{k}{K})^3$ for semantic conditioned motion segments (i.e., $\mathbf{X_p}$ and $\mathbf{X_s}$), and $r=1$ otherwise (i.e., $\mathbf{X_t}$). Note that the value of $r$ for semantically denoised segments is determined by the ratio of total steps $K$ and current denoising step $k$, ensuring transition compatibility early and enriching semantic details progressively. Finally, DDIMScheduler~\cite{song2020ddim} $\mathcal{F}_{\text{D}}$ is utilized to estimate the clean phase latent and the next step phase latent based on the mixed phase latent.

\subsubsection{Motion Inbetweening}
The motion inbetweening task aims to generate an inbetweening motion $\mathbf{X_i}$, which is of a specified length to bridge the gap between two separated motions $[\mathbf{X_p}, \mathbf{X_s}]$.
The pipeline for the task is illustrated in Fig.~\ref{fig:inference_pipeline}. 
The process begins by encoding these segments into latent codes $\mathbf{P}_\mathbf{p}^0$ and $\mathbf{P}_\mathbf{s}^0$ with the ACT‐PAE encoder, then uses TPDMs to guide the generation of inbetweening motion $\mathbf{X_i}$ and two transitioning motions $\mathbf{X_{t_1}}$ and $\mathbf{X_{t_2}}$. An optional SPDM can be incorporated for $\mathbf{X_i}$, modifying the task from unconditional (UMIB) to conditional (CMIB). Additionally, the pipeline demonstrates the flexibility and scalability of the Compositional Phase Diffusion framework by enabling the compositional generation of more motion segments of varying lengths through parallel processing with an increasing number of modules.

\subsubsection{Long-term Motion Generation}
Long-term motion sequence generation extends beyond short-term compositional motion pair generation by producing much longer continuous motion, composed of hundreds or thousands of motion segments. While short-term tasks focus on semantics and transitions within a few segments, long-term generation involves monitoring kinetic dynamics, which can impact motion over extended sequences and potentially disrupt motion realism and physical plausibility. To adapt our compositional motion framework for long-term generation, we can unroll it to process each segment with the [$\text{TPDM}_f$, $\text{SPDM}$,$\text{TPDM}_b$] triplet and denoise them based on semantics and adjacent phase conditions. By rearranging and batching the input for each module, the denoising process of all segments can be done in parallel, making the overall denoising time independent of the number of segments. 

The bidirectional TPDM mechanism in our framework ensures that phase information propagates progressively throughout the sequences, rather than being confined to specific local segments. This mitigates the risk of substantial phase dynamics misalignments between adjacent semantic segments and simplifies the adjustments required by transition segments. Unlike existing methods that struggle with handling substantial differences in motion phase dynamics between segments, which often result in the loss of smooth transitions or semantic alignment, our model continuously refines both motion phase dynamics and text alignment to preserve long-term motion integrity.

\section{Experiments}
\subsection{Implementation and Evaluation Details}\label{sec:implementation}

\subsubsection{Training and Evaluation Dataset}
We use the BABEL-TEACH dataset~\cite{punnakkal2021babel, athanasiou2022teach} for training and evaluation, as it provides annotated subsequence pairs essential for long-term motion generation~\cite{athanasiou2022teach, shafir2024priorMDM, yang2023pcmdm}, facilitating the learning of transitions between subsequences. These annotated pairs are derived from decomposing fine-grained text subsequence annotations from BABEL~\cite{punnakkal2021babel}. For example, a sequence such as [\textit{walk}, \textit{sit down}, \textit{stand up}, \textit{move arms}] is split into pairs like [\textit{walk}, \textit{sit down}], [\textit{sit down}, \textit{stand up}], and [\textit{stand up}, \textit{move arms}]. 

Following the data processing pipeline outlined in recent long-term motion generation~\cite{athanasiou2022teach, shafir2024priorMDM, yang2023pcmdm}, we set the minimum and maximum lengths for each subsequence as 45 and 250 frames, respectively. The pipeline then groups the textually annotated subsequences into pairs. Note that any overlapping offsets identified by the pipeline above are redistributed among the annotated motion subsequences. As a result, the training dataset contains 4370 subsequence pairs, while the testing dataset includes 1582 subsequence pairs. Moreover, we follow PCMDM~\cite{yang2023pcmdm} and priorMDM~\cite{shafir2024priorMDM} to transform the motion data into HumanML3D~\cite{guo2022t2m} format. Initially, the root trajectory is represented by only 4 out of 263 parameters, omitting essential root orientation details. Therefore, we supplement a 6D rotation~\cite{zhou2019rot6d} for the root, increasing the total parameters to $E=269$. Note that all models being compared are trained on the same dataset and representation to ensure a fair comparison.

\textbf{Remark on Dataset.} To the best of our knowledge, BABEL-TEACH is currently the only dataset that provides subsequence pair annotations. Long-term motion generation models, such as TEACH~\cite{athanasiou2022teach}, PCMDM~\cite{yang2023pcmdm}, and our own model, require data samples in the form of continuous subsequence pairs to effectively learn transitions between sequences. Motion datasets annotated in other formats or modalities cannot be efficiently utilized either for training or for fair evaluation of our task.

\begin{figure*}[t]
\centering
\includegraphics[width=0.96\linewidth]{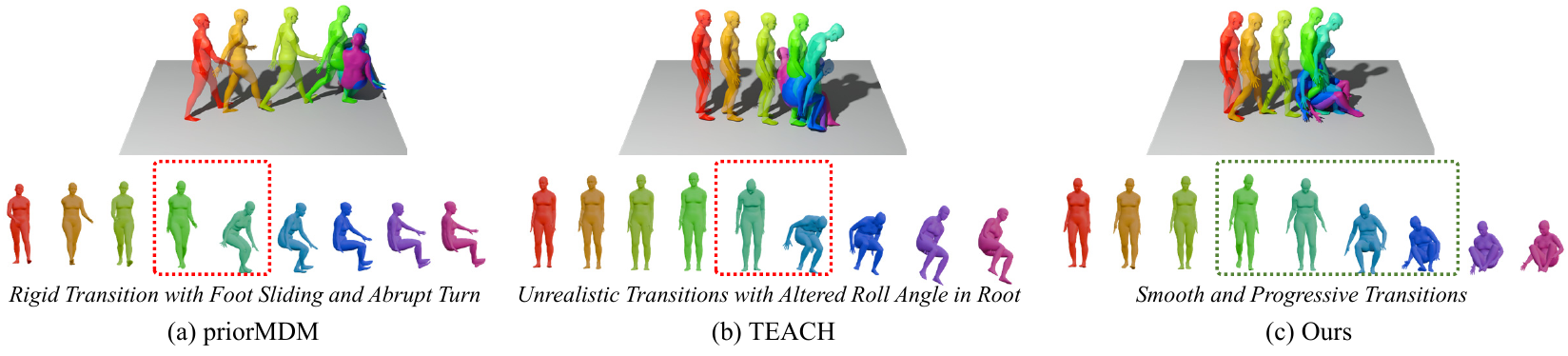}
\vspace{-0.2cm}
\caption{
Compositional motion pair result visualization for [\textit{walk}(2.4s), \textit{sit down}(3.6s)]. The motion frames are colored from red to purple in a rainbow gradient to represent the progression of time. Note that priorMDM exhibits an unrealistic, sudden turn during the sit-down action, which is reflected in its low FID score. TEACH's result includes footing skating in the walk motion. In contrast, our framework generates a fluid walking motion that transitions smoothly into a sit-down action.
}
\vspace{-0.3cm}
\label{fig:t2m_vis}
\end{figure*}

\setlength{\tabcolsep}{4pt}
\begin{table}[t]\small
\caption{Quantitative results for \textbf{Compositional Motion Pair Generation} on the BABEL-TEACH test set. 
\textbf{Bold} and \underline{underline} indicates the \textbf{best} and the \underline{second-best} result.  
}
\label{tab:t2m_qe}
\begin{center}

\scalebox{0.80}{
\begin{tabular}{lcccccc}
\toprule
\textbf{Comp. Motion Pair} & \multicolumn{3}{c}{Motion Realism --- FID$\downarrow$} & \multicolumn{3}{c}{Text Alignment --- MMD$\downarrow$}\\
\cmidrule(lr){2-4}\cmidrule(lr){5-7}
   & Smt. & Trn. & \textbf{Overall} & Smt.  & Trn.  & \textbf{Overall}\\
\midrule

MDM-30~\cite{tevet2023mdm} & 1.084 & 2.526 & 1.146 & 3.793 & \textbf{6.429} & 4.923\\
MLD-30~\cite{chen2023mld} & 13.88 & 15.20 & 14.25 & 8.478 & 7.407 & 7.632\\
TEACH~\cite{athanasiou2022teach}  & \underline{0.941} & 2.375 & 1.041 &  \textbf{3.185} & 7.479 & \underline{4.821}\\
PCMDM~\cite{yang2023pcmdm} &1.056& \underline{1.898}&\underline{0.837}&4.548& 7.323 &5.423\\
priorMDM~\cite{shafir2024priorMDM} & 1.148 & 2.580 & 0.839 & 3.732 &  7.399 & 5.025 \\
\cmidrule(lr){1-7}
Ours & \textbf{0.736} & \textbf{1.807} & \textbf{0.782} & \underline{3.509} & \underline{6.545} & \textbf{4.711}\\
\bottomrule
\end{tabular}}
\end{center}
\vspace{-0.2cm}
\end{table}

\subsubsection{Evaluation Metrics}
We assess the results of \textbf{compositional motion pair generation}, \textbf{long-term motion generation}, and \textbf{conditional motion inbetweening} based on two key aspects: \textit{Fréchet Inception Distance} (FID) for \underline{Motion Realism} and \textit{Multimodal Distance} (MMD) for \underline{Text Alignment}, following the T2M~\cite{guo2022t2m} evaluation protocol utilized in PCMDM~\cite{yang2023pcmdm} and priorMDM~\cite{shafir2024priorMDM} for long-term motion assessment. We exclude \textit{R-precision} (R-prec.) because it overlaps with \textit{Multimodal Distance} (MMD) and omit \textit{Diversity} (Div.) due to its unclear role in evaluating motion performance. For clarity, we segment the generated motions into semantic and transitional parts and evaluate the aforementioned metrics across three groups: 1) Semantic (Smt.), which focuses solely on the semantic segments, 2) Transition (Trn.), which assesses only the transitional segments, and 3) Overall, which evaluates both semantic and transitional segments together. For the contextual alignment of transitional segments, we use the text from both the preceding and succeeding motions, assuming that the transition should retain the semantic information from overlapping segments.

For \textbf{unconditional motion inbetweening}, we assess the \underline{Transition Realism} by using L2 losses and NPSS~\cite{gopalakrishnan2019npss}, as described in \cite{harvey2020RMIB,starke2023MIBPhase}, by comparing them to the ground truth inbetweening motion. Specifically, we focus on L2 losses for joint velocity (\textit{L2-Vel}) and 6D rotation~\cite{zhou2019rot6d} (\textit{L2-Rot6D}) to provide a direct and explicit evaluation of human motion in the HumanML3D~\cite{guo2022t2m} format. 
Additionally, we assess \underline{Transition Smoothness} using root mean squared jerk~\cite{young1997rmsjerk} (RMS-Jerk) over joint rotations. Detailed descriptions of these evaluation metrics will be included in the supplementary material.

\setlength{\tabcolsep}{4pt}
\begin{table}[t]\footnotesize
\caption{Quantitative results for \textbf{Long-term Motion Generation} on the BABEL-TEACH test set with a single extended text sequence of 3,164 texts (302,298 frames, 168 minutes).
% We expand the evaluation metrics by grouping motion segments according to semantic and transition categories, as well as assessing the overall performance when these are combined.
\textbf{Bold} and \underline{underline} indicates the \textbf{best} and the \underline{second-best} result.  
% $\rightarrow$ indicates that the closer to the Input, the better.
}
\label{tab:t2m_long}
\begin{center}

\scalebox{0.8}{
\begin{tabular}{lcccccc}
\toprule
% \toprule
\textbf{Long-term} & \multicolumn{3}{c}{Motion Realism --- FID$\downarrow$} & \multicolumn{3}{c}{Text Alignment --- MMD$\downarrow$}\\
\cmidrule(lr){2-4}\cmidrule(lr){5-7}
   & Smt. & Trn. & \textbf{Overall} & Smt.  & Trn.  & \textbf{Overall}\\
\midrule
MDM-30~\cite{tevet2023mdm} & 1.094 & 2.051 & 1.365 & 3.877 & 6.411 & 4.958\\
MLD-30~\cite{chen2023mld} & 14.36 & 13.44 & 17.02 & 8.423 & 7.407 & 7.690\\
TEACH~\cite{athanasiou2022teach} & \underline{0.785} & 1.645 & 1.780  & \textbf{3.175} & \underline{5.483} & \underline{4.984}\\
PCMDM~\cite{yang2023pcmdm} & 1.068 & \underline{0.934} & \underline{0.876} & 4.188 & 5.721 & 5.156\\
priorMDM~\cite{shafir2024priorMDM} & 2.288 & 1.067 & 1.536 & 4.299 & 5.536 & 5.060\\
\cmidrule(lr){1-7}
Ours & \textbf{0.773} & \textbf{0.909} & \textbf{0.847}  & \underline{3.642} & \textbf{5.389} & \textbf{4.849}\\
\bottomrule
\end{tabular}}
\end{center}
\vspace{-0.3cm}
\end{table}

\subsection{Compositional Motion Generation Performance Evaluation}
\subsubsection{Compositional Motion Pair Generation}
The compositional motion pair experiment follows the setup illustrated on the left in Fig.~\ref{fig:framework}, with the objective of generating motions $\mathbf{X_p}$, $\mathbf{X_t}$, and $\mathbf{X_s}$ based on the corresponding text condition pairs $(C_\mathbf{p},C_\mathbf{s})$. We compare the performance of our method with long motion generation models, including TEACH~\cite{athanasiou2022teach}, PCMDM~\cite{yang2023pcmdm}, and priorMDM~\cite{shafir2024priorMDM}, among with single text-conditioned models MDM-30~\cite{tevet2023mdm} and MLD-30~\cite{chen2023mld} as baselines. In single text-conditioned models, additional frames are generated for preceding and succeeding motions to create a 30-frame overlapping region, which is then blended linearly to form smooth transitions. The evaluation employs the \textit{FID} and \textit{MMD} metrics across three groups: 1) Semantic ($\mathbf{X_p}, \mathbf{X_s}$), 2) Transition ($\mathbf{X_t}$), and 3) Overall ($\mathbf{X_p}, \mathbf{X_t}, \mathbf{X_s}$). The experiment results are summarized in Tab.~\ref{tab:t2m_qe}, showing that our model produces realistic motions and achieves the best overall \textit{FID} and \textit{MMD} scores. Strong overall performance demonstrates the potential to generate high-quality compositional motions with natural transitions. Although TEACH shows strong contextual alignment in semantic segments, its lower \textit{Overall FID} score indicates a compromise in motion quality, leading to motion artifacts such as changes in the root roll angle, as shown in Fig.~\ref{fig:t2m_vis}. 
% Furthermore, while priorMDM achieves the best \textit{Trn. FID}, it produces an unnatural sudden turn artifact, highlighting its limitations despite the metric performance.

\textbf{Remark.} The performance comparison with MDM-30 highlights the effectiveness of SPDM in single-sequence text-to-motion generation. Since only a very short blending window is applied between segments, evaluating \textit{Smt. FID} and \textit{Smt. MMD} for MDM-30 is essentially equivalent to evaluating two independently generated motion sequences, effectively reflecting the single-sequence text-to-motion generation scenario. Therefore, the superior performance of our method on \textit{Smt. FID} and \textit{Smt. MMD} compared to MDM-30 underscores the competitive performance of SPDM in the standalone text-to-motion generation task.

\subsubsection{Long-term Motion Generation}
To assess the long-term motion generation performance, we combine all text conditions from the testing dataset into a single extended text sequence of 3,164 texts, and apply comparison models to generate long motion across 302,298 frames (168 minutes). As shown in Tab.~\ref{tab:t2m_long}, experiment results indicate that our method achieves competitive performance in both motion realism (\textit{Overall FID}) and text alignment (\textit{Overall MMD}) even in long motion generation scenarios. 
This demonstrates the superiority of our phase modeling approach for transition-aware motion generation, compared to other methods that model transitions directly in the raw motion space. 
Among the other compared methods, TEACH continues to struggle with poor transition generation, resulting in a high \textit{Trn. FID}. PriorMDM and PCMDM face challenges in generating realistic semantic segments that align with the text input. 
Note that MDM and MLD show similar performance in this task as in compositional motion pair generation, mainly because the transitions between subsequences are created through blending rather than being generated by machine learning models.

\begin{figure}[t]
\centering
\includegraphics[width=0.98\linewidth]{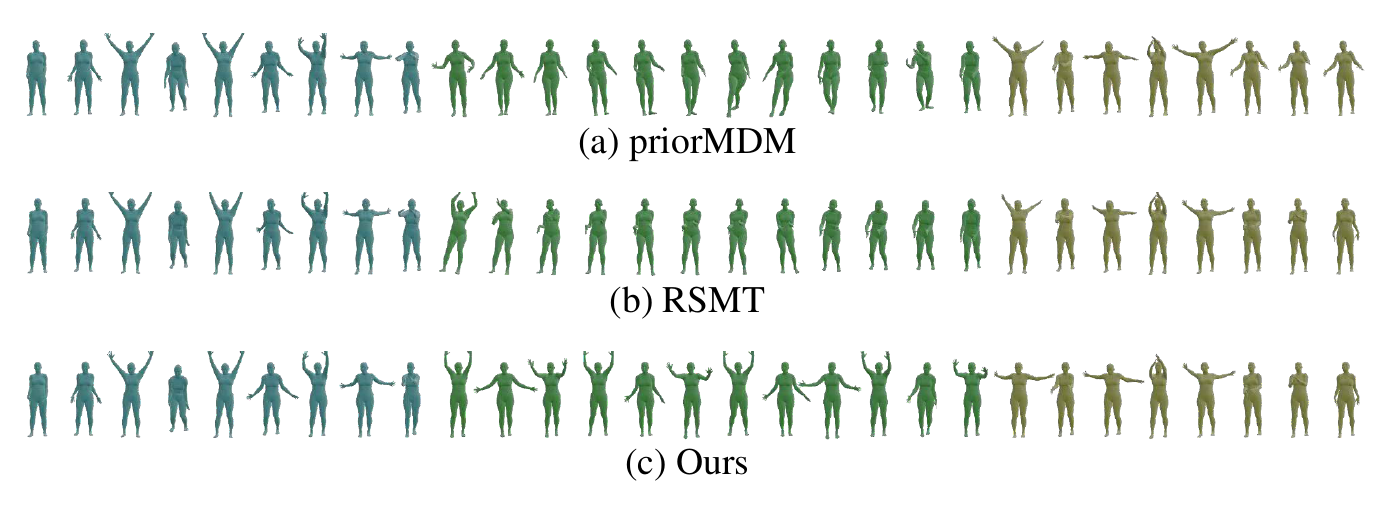}\vspace{-0.46cm}
\caption{
Visualization of the \textbf{UMIB} with 120 transition frames: preceding motion in blue, transitioning motion in green, and succeeding motion in yellow.
}
\label{fig:mib_90}\vspace{-0.15cm}
\end{figure}

\setlength{\tabcolsep}{3.3pt}
\begin{table*}[t]\footnotesize
\caption{Quantitative results for \textbf{Unconditional Motion Inbetweening (UMIB)} on BABEL-TEACH~\cite{athanasiou2022teach} test set. We report the performance under settings of transition lengths at 60, 120, and 180 frames. \textbf{Bold} and \underline{underline} indicates the \textbf{best} and the \underline{second-best} result. }
\vspace{-0.2cm}
\label{tab:mib_qe_umib}
\begin{center}
\scalebox{0.8}{
\begin{tabular}{lcccccccccccccccccc}
\toprule
\textbf{UMIB} & \multicolumn{9}{c}{Transition Realism} & \multicolumn{3}{c}{Transition Smoothness} \\
\cmidrule(lr){2-10}\cmidrule(lr){11-13}
 & \multicolumn{3}{c}{L2-Vel $\downarrow$} &
  \multicolumn{3}{c}{L2-Rot6D $\downarrow$} & \multicolumn{3}{c}{NPSS $\downarrow$} & \multicolumn{3}{c}{RMS-Jerk $\downarrow$}\\
\cmidrule(lr){2-4}\cmidrule(lr){5-7}\cmidrule(lr){8-10}\cmidrule(lr){11-13}
 Length & 60 & 120 & 180 & 60 & 120 & 180 & 60 & 120 & 180 &  60 & 120 & 180 \\
\midrule
RMIB~\cite{harvey2020RMIB} & 0.0353 & 0.0307 & 0.0297 & 0.3510 & 0.3797 &  0.3789   & 2.0908 &5.7334 &  9.1055   &1.9216 & 1.3794 & 1.1009\\
RSMT~\cite{tang2023RSMT} & 0.0345 & 0.0273  &  0.0246  & \underline{0.2151} & 0.2438 &   0.2654   & 0.8552 & 2.8986 &   4.836   & 1.6435 & 1.4670 & 1.3249\\
CMB~\cite{kim2022CMIB} & 0.0200 & 0.0172  & 0.0175  & 0.2194 & \underline{0.2209} & \underline{0.2289}     & \underline{0.4085} & \underline{1.0278} &   \underline{1.6723}   & 2.0658 & 1.9885 & 1.9298\\
MDM~\cite{tevet2023mdm} & 0.0302 & 0.0275  &  0.0319   & 0.2961 & 0.3309 &   0.3099   & 1.0369 & 3.2291 &  4.1440    &1.9283 & 2.1205 & 2.0688\\
priorMDM~\cite{shafir2024priorMDM} & \underline{0.0151} & \underline{0.0141}  &   \underline{0.0144}  & 0.2398 & 0.2455 &   0.2479   & 0.5640 & 1.2107 &   1.7469   & \underline{0.4058} & \underline{0.4495} & \underline{0.2803} \\
\cmidrule(lr){1-13}
Ours & \textbf{0.0101} & \textbf{0.0102} &  \textbf{0.0125}   & \textbf{0.2124} & \textbf{0.2205} &    \textbf{0.2220}  & \textbf{0.3651} & \textbf{0.9296} &    \textbf{1.6308}  & \textbf{0.0963} & \textbf{0.1054} & \textbf{0.1213}\\
\bottomrule
\end{tabular}}
\end{center}
\vspace{-0.35cm}
\end{table*}

\subsection{Motion Inbetweening Performance Evaluation}
The unconditional motion inbetweening (UMIB) experiment follows a setup similar to Fig.~\ref{fig:inference_pipeline}, where a specific number of frames around the transition boundary of testing motion pairs are masked to evaluate various methods for reconstructing the masked motion content. We assess the \textit{L2-Vel}, \textit{L2-Rot6D}, and \textit{NPSS} metrics by comparing the generated segments with the actual masked content, while the \textit{RMS-Jerk} metrics evaluate motion smoothness. The effectiveness of motion inbetweening is analyzed and compared with autoregressive frame prediction methods, RMIB~\cite{harvey2020RMIB} and RSMT~\cite{tang2023RSMT}, as well as interval infilling methods, CMB~\cite{kim2022CMIB}, MDM~\cite{tevet2023mdm}, and priorMDM~\cite{shafir2024priorMDM}, across three different inbetweening length settings, all within the training motion length range of [45, 250].
The results are shown in Tab.~\ref{tab:mib_qe_umib}, demonstrating the superior performance of our proposed framework across all inbetweening length settings. Additionally, Fig.~\ref{fig:mib_90} illustrates the smoothness and realism of our generated results, as reflected in the metric values. In contrast, the results generated by priorMDM tends to exhibit hyperactivity by producing random motion content unrelated to the adjacent motion, which negatively impacts the overall inbetweening performance. 
Lastly, RMST results reveal a failure to connect the succeeding motions. This highlights the limitations of autoregressive processing with fixed-window phase latents, which also justifies both priorMDM and our method for managing variable-length motion as a cohesive entity.

In addition to the UMIB experiment, we also conducted the conditional motion inbetweening (CMIB) experiment to assess the effectiveness of conditioning the inbetweening region with text context. The results of this experiment are detailed in the Appendix.

\subsection{Ablation Studies and User Studies}\label{sec:ablations}
We assess the effects of our proposed modules and recommended hyperparameters on compositional motion generation and motion inbetweening tasks. Firstly, the integration of \textit{frame-level tokens} within SPDM and TPDM significantly enhances their performance in denoising \textit{param-level tokens}. Secondly, we assess the phase mixing parameter setting, revealing that $r = (\frac{k}{K})^3$ is optimal for semantic conditional scenarios, while $r = 1$ is best for unconditional scenarios. Moreover, in the user study, our approach attains the highest scores for motion realism and smoothness. Further details of the ablation studies will be provided in the supplementary material.

\section{Conclusion and Future Work}\label{sec:conclusions}
We present the Transitional Phase Diffusion Module (TPDM) and the Semantic Phase Diffusion Module (SPDM), which operate within the periodic latent space generated by the Action-Centric Periodic Autoencoder. These modules inject semantic guidance and neighbouring phase information into the motion denoising process, enabling the generation of semantically meaningful motion clips with smooth transitions.
The proposed Compositional Phase Diffusion pipeline, which incorporates both the TPDM and SPDM modules, can be adapted for compositional motion generation and motion inbetweening tasks. Its flexibility to handle multiple motion segments simultaneously enhances its capability to tackle complex motion sequencing tasks. Extensive experiments and evaluations have showcased the framework's effectiveness in compositional motion generation and motion inbetweening tasks. Further exploration of these frameworks holds promise for the development of advanced motion-generation techniques in the future. 
As our framework applies compositional diffusion in motion generation using a basic phase mixing technique, potential performance improvement may be achievable by incorporating advanced methods like score-based or potential-based diffusion. Additionally, incorporating learnable parameters or an adaptive mechanism for phase mixing could further enhance results.
However, implementing such features is challenging and requires more detailed data modelling and complex architectures. Future research will focus on adjusting the architecture and data representation to incorporate these advanced diffusion techniques.

\bibliographystyle{unsrt}
\bibliography{myRefs}

\begin{thebibliography}{10}

\bibitem{vaswani2017transformer}
Ashish Vaswani, Noam Shazeer, Niki Parmar, Jakob Uszkoreit, Llion Jones, Aidan~N Gomez, {\L}ukasz Kaiser, and Illia Polosukhin.
\newblock Attention is all you need.
\newblock {\em Advances in neural information processing systems}, 30, 2017.

\bibitem{ho2020ddpm}
Jonathan Ho, Ajay Jain, and Pieter Abbeel.
\newblock Denoising diffusion probabilistic models.
\newblock {\em Advances in neural information processing systems}, 33:6840--6851, 2020.

\bibitem{song2020ddim}
Jiaming Song, Chenlin Meng, and Stefano Ermon.
\newblock Denoising diffusion implicit models.
\newblock {\em arXiv preprint arXiv:2010.02502}, 2020.

\bibitem{punnakkal2021babel}
Abhinanda~R Punnakkal, Arjun Chandrasekaran, Nikos Athanasiou, Alejandra Quiros-Ramirez, and Michael~J Black.
\newblock Babel: Bodies, action and behavior with english labels.
\newblock In {\em Proceedings of the IEEE/CVF Conference on Computer Vision and Pattern Recognition}, pages 722--731, 2021.

\bibitem{jiang2022dual}
Junkun Jiang, Jie Chen, and Yike Guo.
\newblock A dual-masked auto-encoder for robust motion capture with spatial-temporal skeletal token completion.
\newblock In {\em Proceedings of the 30th ACM International Conference on Multimedia}, pages 5123--5131, 2022.

\bibitem{guo2022t2m}
Chuan Guo, Shihao Zou, Xinxin Zuo, Sen Wang, Wei Ji, Xingyu Li, and Li~Cheng.
\newblock Generating diverse and natural 3d human motions from text.
\newblock In {\em Proceedings of the IEEE/CVF Conference on Computer Vision and Pattern Recognition}, pages 5152--5161, 2022.

\bibitem{jiang2025every}
Junkun Jiang, Jie Chen, Ho~Yin Au, Mingyuan Chen, Wei Xue, and Yike Guo.
\newblock Every angle is worth a second glance: Mining kinematic skeletal structures from multi-view joint cloud.
\newblock {\em IEEE Transactions on Visualization and Computer Graphics}, 2025.

\bibitem{tevet2023mdm}
Guy Tevet, Sigal Raab, Brian Gordon, Yoni Shafir, Daniel Cohen-or, and Amit~Haim Bermano.
\newblock Human motion diffusion model.
\newblock In {\em The Eleventh International Conference on Learning Representations}, 2023.

\bibitem{guo2022tm2t}
Chuan Guo, Xinxin Zuo, Sen Wang, and Li~Cheng.
\newblock Tm2t: Stochastic and tokenized modeling for the reciprocal generation of 3d human motions and texts.
\newblock In {\em European Conference on Computer Vision}, pages 580--597. Springer, 2022.

\bibitem{kim2023flame}
Jihoon Kim, Jiseob Kim, and Sungjoon Choi.
\newblock Flame: Free-form language-based motion synthesis \& editing.
\newblock In {\em Proceedings of the AAAI Conference on Artificial Intelligence}, volume~37, pages 8255--8263, 2023.

\bibitem{zhang2024motiondiffuse}
Mingyuan Zhang, Zhongang Cai, Liang Pan, Fangzhou Hong, Xinying Guo, Lei Yang, and Ziwei Liu.
\newblock Motiondiffuse: Text-driven human motion generation with diffusion model.
\newblock {\em IEEE Transactions on Pattern Analysis and Machine Intelligence}, 2024.

\bibitem{athanasiou2022teach}
Nikos Athanasiou, Mathis Petrovich, Michael~J Black, and G{\"u}l Varol.
\newblock Teach: Temporal action composition for 3d humans.
\newblock In {\em 2022 International Conference on 3D Vision (3DV)}, pages 414--423. IEEE, 2022.

\bibitem{shafir2024priorMDM}
Yoni Shafir, Guy Tevet, Roy Kapon, and Amit~Haim Bermano.
\newblock Human motion diffusion as a generative prior.
\newblock In {\em The Twelfth International Conference on Learning Representations}, 2024.

\bibitem{yang2023pcmdm}
Zhao Yang, Bing Su, and Ji-Rong Wen.
\newblock Synthesizing long-term human motions with diffusion models via coherent sampling.
\newblock In {\em Proceedings of the 31st ACM International Conference on Multimedia}, pages 3954--3964, 2023.

\bibitem{starke2022deepphase}
Sebastian Starke, Ian Mason, and Taku Komura.
\newblock Deepphase: Periodic autoencoders for learning motion phase manifolds.
\newblock {\em ACM Transactions on Graphics (TOG)}, 41(4):1--13, 2022.

\bibitem{holden2017PFNN}
Daniel Holden, Taku Komura, and Jun Saito.
\newblock Phase-functioned neural networks for character control.
\newblock {\em ACM Transactions on Graphics (TOG)}, 36(4):1--13, 2017.

\bibitem{starke2020localphase}
Sebastian Starke, Yiwei Zhao, Taku Komura, and Kazi Zaman.
\newblock Local motion phases for learning multi-contact character movements.
\newblock {\em ACM Transactions on Graphics (TOG)}, 39(4):54--1, 2020.

\bibitem{mason2022stylelocalphase}
Ian Mason, Sebastian Starke, and Taku Komura.
\newblock Real-time style modelling of human locomotion via feature-wise transformations and local motion phases.
\newblock {\em Proceedings of the ACM on Computer Graphics and Interactive Techniques}, 5(1):1--18, 2022.

\bibitem{starke2023MIBPhase}
Paul Starke, Sebastian Starke, Taku Komura, and Frank Steinicke.
\newblock Motion in-betweening with phase manifolds.
\newblock {\em Proceedings of the ACM on Computer Graphics and Interactive Techniques}, 6(3):1--17, 2023.

\bibitem{tang2023RSMT}
Xiangjun Tang, Linjun Wu, He~Wang, Bo~Hu, Xu~Gong, Yuchen Liao, Songnan Li, Qilong Kou, and Xiaogang Jin.
\newblock Rsmt: Real-time stylized motion transition for characters.
\newblock In {\em ACM SIGGRAPH 2023 Conference Proceedings}, pages 1--10, 2023.

\bibitem{wan2023diffusionphase}
Weilin Wan, Yiming Huang, Shutong Wu, Taku Komura, Wenping Wang, Dinesh Jayaraman, and Lingjie Liu.
\newblock Diffusionphase: Motion diffusion in frequency domain.
\newblock {\em arXiv preprint arXiv:2312.04036}, 2023.

\bibitem{chen2023mld}
Xin Chen, Biao Jiang, Wen Liu, Zilong Huang, Bin Fu, Tao Chen, and Gang Yu.
\newblock Executing your commands via motion diffusion in latent space.
\newblock In {\em Proceedings of the IEEE/CVF Conference on Computer Vision and Pattern Recognition}, pages 18000--18010, 2023.

\bibitem{radford2021clip}
Alec Radford, Jong~Wook Kim, Chris Hallacy, Aditya Ramesh, Gabriel Goh, Sandhini Agarwal, Girish Sastry, Amanda Askell, Pamela Mishkin, Jack Clark, et~al.
\newblock Learning transferable visual models from natural language supervision.
\newblock In {\em International conference on machine learning}, pages 8748--8763. PMLR, 2021.

\bibitem{yuan2023physdiff}
Ye~Yuan, Jiaming Song, Umar Iqbal, Arash Vahdat, and Jan Kautz.
\newblock Physdiff: Physics-guided human motion diffusion model.
\newblock In {\em Proceedings of the IEEE/CVF international conference on computer vision}, pages 16010--16021, 2023.

\bibitem{karunratanakul2023gmd}
Korrawe Karunratanakul, Konpat Preechakul, Supasorn Suwajanakorn, and Siyu Tang.
\newblock Guided motion diffusion for controllable human motion synthesis.
\newblock In {\em Proceedings of the IEEE/CVF International Conference on Computer Vision}, pages 2151--2162, 2023.

\bibitem{harvey2018RTN}
F{\'e}lix~G Harvey and Christopher Pal.
\newblock Recurrent transition networks for character locomotion.
\newblock In {\em SIGGRAPH Asia 2018 Technical Briefs}, pages 1--4. 2018.

\bibitem{harvey2020RMIB}
F{\'e}lix~G Harvey, Mike Yurick, Derek Nowrouzezahrai, and Christopher Pal.
\newblock Robust motion in-betweening.
\newblock {\em ACM Transactions on Graphics (TOG)}, 39(4):60--1, 2020.

\bibitem{kaufmann2020CMIF}
Manuel Kaufmann, Emre Aksan, Jie Song, Fabrizio Pece, Remo Ziegler, and Otmar Hilliges.
\newblock Convolutional autoencoders for human motion infilling.
\newblock In {\em 2020 International Conference on 3D Vision (3DV)}, pages 918--927. IEEE, 2020.

\bibitem{hernandez2019STinpaint}
Alejandro Hernandez, Jurgen Gall, and Francesc Moreno-Noguer.
\newblock Human motion prediction via spatio-temporal inpainting.
\newblock In {\em Proceedings of the IEEE/CVF International Conference on Computer Vision}, pages 7134--7143, 2019.

\bibitem{kim2022CMIB}
Jihoon Kim, Taehyun Byun, Seungyoun Shin, Jungdam Won, and Sungjoon Choi.
\newblock Conditional motion in-betweening.
\newblock {\em Pattern Recognition}, 132:108894, 2022.

\bibitem{chi2024m2d2m}
Seunggeun Chi, Hyung-gun Chi, Hengbo Ma, Nakul Agarwal, Faizan Siddiqui, Karthik Ramani, and Kwonjoon Lee.
\newblock M2d2m: Multi-motion generation from text with discrete diffusion models.
\newblock In {\em European Conference on Computer Vision}, pages 18--36. Springer, 2024.

\bibitem{zhang2024infinimotion}
Zeyu Zhang, Akide Liu, Qi~Chen, Feng Chen, Ian Reid, Richard Hartley, Bohan Zhuang, and Hao Tang.
\newblock Infinimotion: Mamba boosts memory in transformer for arbitrary long motion generation.
\newblock {\em arXiv preprint arXiv:2407.10061}, 2024.

\bibitem{kovar2008motiongraph}
Lucas Kovar, Michael Gleicher, and Fr{\'e}d{\'e}ric Pighin.
\newblock Motion graphs.
\newblock {\em ACM Transactions on Graphics}, 21(3):473--482, 2002.

\bibitem{min2012motiongraphpp}
Jianyuan Min and Jinxiang Chai.
\newblock Motion graphs++: a compact generative model for semantic motion analysis and synthesis.
\newblock {\em ACM Transactions on Graphics}, 31(6):1--12, 2012.

\bibitem{au2022choreograph}
Ho~Yin Au, Jie Chen, Junkun Jiang, and Yike Guo.
\newblock Choreograph: Music-conditioned automatic dance choreography over a style and tempo consistent dynamic graph.
\newblock In {\em Proceedings of the 30th ACM International Conference on Multimedia}, pages 3917--3925, 2022.

\bibitem{au2024rechoreonet}
Ho~Yin Au, Jie Chen, Junkun Jiang, and Yike Guo.
\newblock Rechoreonet: Repertoire-based dance re-choreography with music-conditioned temporal and style clues.
\newblock {\em Machine Intelligence Research}, pages 1--11, 2024.

\bibitem{petrovich2021actor}
Mathis Petrovich, Michael~J Black, and G{\"u}l Varol.
\newblock Action-conditioned 3d human motion synthesis with transformer vae.
\newblock In {\em Proceedings of the IEEE/CVF International Conference on Computer Vision}, pages 10985--10995, 2021.

\bibitem{zhou2019rot6d}
Yi~Zhou, Connelly Barnes, Jingwan Lu, Jimei Yang, and Hao Li.
\newblock On the continuity of rotation representations in neural networks.
\newblock In {\em Proceedings of the IEEE/CVF conference on computer vision and pattern recognition}, pages 5745--5753, 2019.

\bibitem{gopalakrishnan2019npss}
Anand Gopalakrishnan, Ankur Mali, Dan Kifer, Lee Giles, and Alexander~G Ororbia.
\newblock A neural temporal model for human motion prediction.
\newblock In {\em Proceedings of the IEEE/CVF Conference on Computer Vision and Pattern Recognition}, pages 12116--12125, 2019.

\bibitem{young1997rmsjerk}
Raymond~P Young and Ronald~G Marteniuk.
\newblock Acquisition of a multi-articular kicking task: Jerk analysis demonstrates movements do not become smoother with learning.
\newblock {\em Human Movement Science}, 16(5):677--701, 1997.

\end{thebibliography}

\newpage
\appendix
\section{Technical Details of Compositional Phase Diffusion}\label{sec:appendix}
\subsection{Compositional Motion Generation Algorithm}
Algorithm implementation details of the phase diffusion pipeline for the compositional motion generation task in Sec. \ref{sec:compositional}. The detail is also illustrated in Fig. \ref{fig:framework}.
\begin{algorithm}[ht]\small
\caption{Compositional Phase Diffusion on generating sequence composed by 2 semantic conditioned segments}\label{alg:diffusion}
\begin{algorithmic}
\State \textbf{Require:} forward TPDM $\mathcal{F}_{\text{T}_f}(\cdot)$, backward TPDM $\mathcal{F}_{\text{T}_b}(\cdot)$, SPDM $\mathcal{F}_{\text{S}}(\cdot)$, DiffusionScheduler $\mathcal{F}_{\text{D}}(\cdot)$
% \\ \textcolor{gray}{\# initialize phase latents for $\mathbf{X_p}$, $\mathbf{X_s}$, $\mathbf{X_t}$}
\State $\mathbf{P}_\mathbf{p}^K, \mathbf{P}_\mathbf{s}^K, \mathbf{P}_\mathbf{t}^K \sim \mathbb{N}(\mathbf{0},\mathbf{I})$ \Comment{sample latent for $\mathbf{X_p}$, $\mathbf{X_s}$, $\mathbf{X_t}$}
\State $\mathbf{P}_\mathbf{p}^0, \mathbf{P}_\mathbf{s}^0, \mathbf{P}_\mathbf{t}^0 \gets \mathbf{P}_\mathbf{p}^K, \mathbf{P}_\mathbf{s}^K, \mathbf{P}_\mathbf{t}^K$ \Comment{clean latent placeholder}
\For{$k$ from $K$ to $1$}
\State \textcolor{gray}{ \# predict phase latent noise using SPDM and TPDMs}
\State $\mathbf{P}_{\mathbf{p}c}^0 \gets \mathcal{F}_{\text{S}}(k, C_\mathbf{p}, \mathbf{P}_\mathbf{p}^k), \mathbf{P}_{\mathbf{p}b}^0 \gets \mathcal{F}_{\text{T}_b}(k, \mathbf{P}_\mathbf{p}^k, \mathbf{P}_\mathbf{t}^0)$
\State $\mathbf{P}_{\mathbf{t}f}^0 \gets \mathcal{F}_{\text{T}_f}(k, \mathbf{P}_\mathbf{t}^k, \mathbf{P}_\mathbf{p}^0), \mathbf{P}_{\mathbf{t}b}^0 \gets \mathcal{F}_{\text{T}_b}(k, \mathbf{P}_\mathbf{t}^k, \mathbf{P}_\mathbf{s}^0)$
\State $\mathbf{P}_{\mathbf{s}f}^0 \gets \mathcal{F}_{\text{T}_f}(k, \mathbf{P}_\mathbf{s}^k, \mathbf{P}_\mathbf{t}^0), \mathbf{P}_{\mathbf{s}c}^0 \gets \mathcal{F}_{\text{S}}(k, C_\mathbf{s}, \mathbf{P}_\mathbf{s}^k)$
\State \textcolor{gray}{\# perform phase mixing as in Eq.~\ref{eq:phase_mix}}
\State $\mathbf{P}_\mathbf{p}^0 \gets (\frac{k}{K})^3 \mathbf{P}_{\mathbf{p}b}^0 + (1-(\frac{k}{K})^3) \mathbf{P}_{\mathbf{p}c}^0$  \Comment{$r_\mathbf{p} = (\frac{k}{K})^3$}
\State $\mathbf{P}_\mathbf{t}^0 \gets \frac{\mathbf{P}_{\mathbf{t}f}^0 + \mathbf{P}_{\mathbf{t}b}^0}{2}$ \Comment{$r_\mathbf{t} = 1$}
\State $\mathbf{P}_\mathbf{s}^0 \gets (\frac{k}{K})^3 \mathbf{P}_{\mathbf{s}f}^0 + (1-(\frac{k}{K})^3) \mathbf{P}_{\mathbf{s}c}^0$  \Comment{$r_\mathbf{s} = (\frac{k}{K})^3$}
\State \textcolor{gray}{\# estimate phase latent at both step $k-1$ and step $0$}

\State $\mathbf{P}_\mathbf{p}^{k-1}, \mathbf{P}_\mathbf{p}^0 \gets \mathcal{F}_{\text{D}}(\mathbf{P}_\mathbf{p}^0, \mathbf{P}_\mathbf{p}^k, k-1)$
\State $\mathbf{P}_\mathbf{t}^{k-1}, \mathbf{P}_\mathbf{t}^0  \gets \mathcal{F}_{\text{D}}(\mathbf{P}_\mathbf{t}^0, \mathbf{P}_\mathbf{t}^k, k-1)$
\State $\mathbf{P}_\mathbf{s}^{k-1}, \mathbf{P}_\mathbf{s}^0 \gets \mathcal{F}_{\text{D}}(\mathbf{P}_\mathbf{s}^0, \mathbf{P}_\mathbf{s}^k, k-1)$
\EndFor
\State \Return $\mathbf{P}_\mathbf{p}^0,\mathbf{P}_\mathbf{t}^0,\mathbf{P}_\mathbf{s}^0$
\end{algorithmic}
\end{algorithm}

\subsection{Adjustment to $T$ and $PE$}
As discussed in Sec.~\ref{sec:actpae}, we have refined the sinusoidal positional embedding $PE$ and the time window $T$ to support motion autoencoding with variable lengths. 

Positional embedding $PE$ is crucial for accurately representing time progression in motion encoding and generation. Traditional sinusoidal positional embeddings $\textbf{PE}$ only reflect the time progression signal from the leading frame of motion $\textit{leading}$ frame of motion $\mathbf{X}$. In contrast, our composite positional embedding $\textbf{Comp-PE}$ creates duplicates of the positional embedding shifted to the $\textit{middle}$ and $\textit{ending}$ frames. These duplicates are stacked channel-wise, enhancing the model’s awareness of sequential action progression from three key locations and improving semantic understanding.

On the other hand, the time window $T \in \mathbb{R}^{N \times Q}$ is essential for transforming the fixed-size parameters $\mathbf{F},\mathbf{A},\mathbf{B},\mathbf{S}$ into the variable length periodic signal $\mathbf{Q}$ following Equation~\ref{eq:phase_reparam} in the main paper.
In traditional PAEs~\cite{starke2022deepphase, wan2023diffusionphase}, $T$ is defined as a fixed‐length time window ranging from $-1$ to $1$ across $121$ frames. When extending $T$ to adapt to variable lengths, two variants arise: $(\textbf{normT})$ parameterizes the time window from $-1$ to $1$ over the frame count $N$ to correspond with normalized action progression, and $(\textbf{frameT})$ parameterizes the time window from $-\frac{N}{2}$ to $\frac{N}{2}$ to align with the actual time duration.
In this work, we employ a mixed linear parameterization $(\textbf{mixT})$, where $\frac{Q}{2}$ channels within the time window $T$ are parameterized using $\textbf{frameT}$, while the other half is parameterized using $\textbf{normT}$. 

Note that the time window parameterizations are implemented piecewise to address the length imbalance in the transitioning motion $\mathbf{X_t}$\footnote{$\mathbf{X_t}$ represents the segment covering the second half of $\mathbf{X_p}$ and the first half of $\mathbf{X_s}$. For example, if $\mathbf{X_p}$ is 2 seconds and $\mathbf{X_s}$ is 8 seconds, $\mathbf{X_t}$ will span 5 seconds, covering the last 1 second of $\mathbf{X_p}$ and the first 4 seconds of $\mathbf{X_s}$. Note that the middle frame of $\mathbf{X_t}$  is defined at the transition boundary; when $\mathbf{X_p}$ and $\mathbf{X_s}$ have unequal lengths, this middle frame is offset from the center of $\mathbf{X_t}$.}. Using the settings in Fig.~\ref{fig:framework} of the main paper as an example, $\textbf{normT}$ parameterize the $\mathbf{X_t}$ left frame range $[\mathbf{t_{tl}}, \mathbf{t_{tm}}]$ as $[-1,0]$, and the $\mathbf{X_t}$ right frame range $[\mathbf{t_{tm}}, \mathbf{t_{te}}]$ as $[0,1]$. Similarly, $\textbf{frameT}$ parameterize the $\mathbf{X_t}$ right frame range $[\mathbf{t_{pm}}, \mathbf{t_{te}}]$ as $[0,\mathbf{t_{te}}-\mathbf{t_{tm}}]$ to align with the actual time duration.

\subsection{Details of SPDM and TPDM}
As discussed in Sec.~\ref{sec:spdm} and Sec.~\ref{sec:tpdm} in the main paper, our diffusion models are designed as $\epsilon$-model~\cite{karunratanakul2023gmd}, which is trained using the $\ell_1$ loss (formulated as $||\,||_1 $) to the diffusion noise $[\epsilon_\mathbf{p}, \epsilon_\mathbf{t}, \epsilon_\mathbf{s}]$ which was scheduled to diffuse the clean phase latent of the motion segments to $[\mathbf{P}_\mathbf{p}^k, \mathbf{P}_\mathbf{t}^k, \mathbf{P}_\mathbf{s}^k]$ respectively. Note that the estimation of diffused latent at the next diffusion step (prev\_sample) and the clean phase latent (pred\_original\_sample) is also supported by the commonly used DDIM diffusion scheduler~\cite{song2020ddim}: $\mathbf{P}_\mathbf{p}^{k-1}, \mathbf{P}_\mathbf{p}^0 \gets \mathcal{F}_{\text{D}}(\mathbf{P}_\mathbf{p}^0, \mathbf{P}_\mathbf{p}^k, k-1)$.

SPDM is trained using semantically annotated motion segments in the BABEL-TEACH~\cite{athanasiou2022teach} training dataset, which are the preceding motion $\mathbf{X_p}$ and succeeding motion $\mathbf{X_s}$ in each motion subsequence pair, each associated with text annotations $C_\mathbf{p}$ and $C_\mathbf{s}$. The training loss of SPDM on each subsequence pair is illustrated as follows:
\begin{equation}
\nonumber
\scalebox{0.87}{$
\mathcal{L}_\text{S} = ||\mathcal{F}_{\text{S}}(k, C_\mathbf{p}, \mathbf{P}_\mathbf{p}^k) - \epsilon_\mathbf{p}||_1 +||\mathcal{F}_{\text{S}}(k, C_\mathbf{s}, \mathbf{P}_\mathbf{s}^k) - \epsilon_\mathbf{s}||_1 .
$}
\end{equation}

On the other hand, TPDM is trained based on the neighbouring information in each motion subsequence pair. Specifically, we can obtain 2 transitional segment pair $(\mathbf{X_p}, \mathbf{X_t})$, $(\mathbf{X_t}, \mathbf{X_s})$ for each motion data tuple $(\mathbf{X_p}, \mathbf{X_t}, \mathbf{X_s})$. Then, $\text{TPDM}_f$ and $\text{TPDM}_b$ are trained on each transitional segment pair to denoise motion phase parameters using neighbouring phase information from either the forward or backward direction. The training loss of TPDMs on each subsequence pair are illustrated as follows:
\begin{equation}
\nonumber
\scalebox{0.87}{$
\begin{aligned}
\mathcal{L}_{\text{T}_f} = ||\mathcal{F}_{\text{T}_f}(k, \mathbf{P}_\mathbf{t}^k, \mathbf{P}_\mathbf{p}^0) - \epsilon_\mathbf{t}||_1 &  + ||\mathcal{F}_{\text{T}_f}(k, \mathbf{P}_\mathbf{s}^k, \mathbf{P}_\mathbf{t}^0) - \epsilon_\mathbf{s}||_1 ,\\
\mathcal{L}_{\text{T}_b} = ||\mathcal{F}_{\text{T}_b}(k, \mathbf{P}_\mathbf{p}^k, \mathbf{P}_\mathbf{t}^0) - \epsilon_\mathbf{p}||_1 & + ||\mathcal{F}_{\text{T}_b}(k, \mathbf{P}_\mathbf{t}^k, \mathbf{P}_\mathbf{s}^0) - \epsilon_\mathbf{t}||_1 .
\end{aligned}$}
\end{equation}

\subsection{Implementation Details}
 We apply the emphasis projection with $c=15$, as demonstrated in GMD~\cite{karunratanakul2023gmd}, to incorporate root trajectory information into the motion representation. Also, our models are designed based on phase latent size $Q=512$, which serves as both the latent dimension for all diffusion modules and the number of periodic signals in ACT‐PAE. For the diffusion step setting in SPDM and TPDM, DDIM~\cite{song2020ddim} is utilized for 1000 training steps and 100 inference steps.
\section{Conditional Motion Inbetweening Evaluation}

\begin{figure}[ht]
\centering
\includegraphics[width=0.98\linewidth]{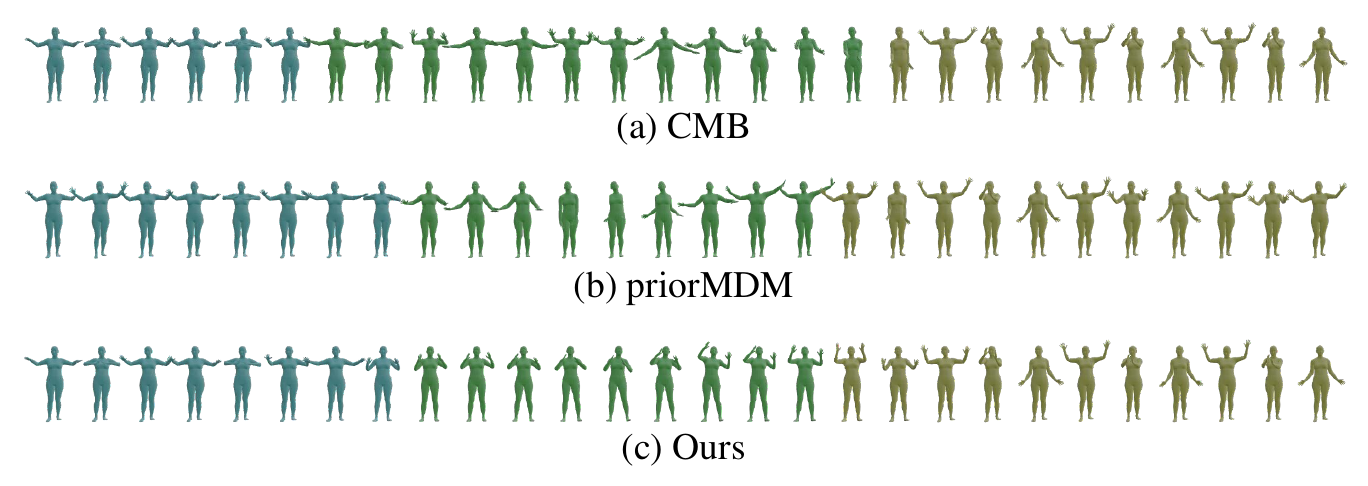}
% \vspace{-0.26cm}
\caption{
Visualization of the \textbf{CMIB} with 120 transition boundary frames conditioned with \textit{bend arms up}: preceding motion in blue, transitioning motion in green, and succeeding motion in yellow.
}
% \vspace{-0.1cm}
\label{fig:cmib_60}
\end{figure}

\setlength{\tabcolsep}{6pt}
\begin{table}[ht]\footnotesize
\caption{Quantitative results for \textbf{Conditional Motion Inbetweening (CMIB)} on the BABEL-TEACH~\cite{athanasiou2022teach} test set. We report the performance of various methods under the settings of transition lengths at 60, 120, and 180 frames. \textbf{Bold} and \underline{underline} indicates the \textbf{best} and the \underline{second-best} result.}
\label{tab:mib_qe_cmib}
\begin{center}
\scalebox{0.8}{
\begin{tabular}{lcccccccc}
\toprule
% \toprule & 
\textbf{CMIB} & \multicolumn{3}{c}{Motion Realism } & \multicolumn{3}{c}{Text Alignment} \\
\cmidrule(lr){2-4}\cmidrule(lr){5-7}
 & \multicolumn{3}{c}{Smt. FID $\downarrow$} & \multicolumn{3}{c}{MMD $\downarrow$}\\
\cmidrule(lr){2-4}\cmidrule(lr){5-7}
Length  & 60 &  120 & 180 & 60 & 120 & 180 \\
\midrule
CMB &  \underline{0.693} & \underline{1.382} &  2.765 &7.420 & \underline{7.544} & 7.561\\
MDM & 0.694 & 1.482 & \underline{2.626} &\underline{7.411} &7.609& 7.658\\
priorMDM~\cite{shafir2024priorMDM} & 1.613 &1.392 & 5.699 & 7.761 & 8.075 & \underline{7.544} \\
\cmidrule(lr){1-7}
Ours & \textbf{0.389}  & \textbf{0.679} & \textbf{2.152}&  \textbf{7.213}  & \textbf{6.871} & \textbf{7.206}\\
\bottomrule
% \bottomrule
\end{tabular}}
\end{center}
% \vspace{-0.4cm}
\end{table}

As shown in Fig.~\ref{fig:inference_pipeline}, our framework can integrate SPDM into the denoising process of inbetweening segments to enable conditional motion inbetweening. We use the same testing setup as in unconditional motion inbetweening, focusing the evaluation on the inbetweening region. We evaluate our framework against CMB~\cite{kim2022CMIB}, MDM~\cite{tevet2023mdm}, and priorMDM~\cite{shafir2024priorMDM}. The results, presented in Tab.~\ref{tab:mib_qe_cmib}, demonstrate that our method excels at producing natural inbetweening motion while adapting to input text semantics. As illustrated in Fig.~\ref{fig:cmib_60}, our framework creates smooth inbetweening motion that corresponds well with the input text condition $\textit{bend arms up}$. In contrast, both priorMDM and CMB show hyperactivity, resulting in abrupt inbetweening motion that does not align with the text.

\section{Impact Statements}
\label{appendix:impact}
The exploration and application of phase latent spaces in this work contribute to the advancement of deep learning by offering new methodologies for signal processing and multimedia generation. It has no negative impact on society as the focus is on technological improvement rather than datasets that could be sensitive or have privacy implications.

\end{document}